\documentclass[12pt,preprint]{aastex}
\usepackage{lscape}

\newcommand{\um}{$\mu$m~}
\newcommand{\ums}{$\mu$m}

\shorttitle{Spectra Of AGN}
\shortauthors{Weedman et al.}

\begin{document}

\title{Mid-Infrared Spectra of Classical AGN Observed with the Spitzer Space Telescope}

\author{D. W. Weedman\altaffilmark{1}, Lei Hao\altaffilmark{1}, S. J. U. Higdon\altaffilmark{1}, D. Devost\altaffilmark{1}, Yanling Wu\altaffilmark{1}, V. Charmandaris\altaffilmark{2}, B. Brandl\altaffilmark{3}, E. Bass\altaffilmark{1}, J. R. Houck\altaffilmark{1}}

\altaffiltext{1}{Astronomy Department, Cornell University, Ithaca, NY 14853; dweedman@astro.cornell.edu}
\altaffiltext{2}{University of Crete, Department of Physics, P.O. Box 2208, GR-71003, Heraklion, Greece}
\altaffiltext{3}{Leiden Observatory, 2300 RA Leiden, The Netherlands}

\begin{abstract}

 Full low resolution (65 $<$R$<$ 130) and high resolution (R $\sim$ 600) spectra between 5\,$\mu$m and 37\,$\mu$m obtained with the Infrared Spectrograph (IRS) on the Spitzer Space Telescope are presented for eight classical active galactic nuclei (AGN) which have been extensively studied previously. Spectra of these AGN are presented as comparison standards for the many objects, including sources at high redshift, which are being observed spectroscopically in the mid-infrared for the first time using the IRS. The AGN are NGC4151, Markarian 3, I Zwicky 1, NGC 1275, Centaurus A, NGC 7469, Markarian 231, and NGC 3079. These sources are used to demonstrate the range of infrared spectra encountered in objects which have widely different classification criteria at other wavelengths but which unquestionably contain AGN.  Overall spectral characteristics - including continuum shape, nebular emission lines, silicate absorption and emission features, and PAH emission features - are considered to understand how spectral classifications based on mid-infrared spectra relate to those previously derived from optical spectra.   The AGN are also compared to the same parameters for starburst galaxies such as NGC 7714 and the compact, low metallicity starburst SBS 0335-052 previously observed with the IRS. Results confirm the much lower strengths of PAH emission features in AGN, but there are no spectral parameters in this sample which unambiguously distinguish AGN and starbursts based only on the slopes of the continuous spectra.

\end{abstract}



\section{Introduction}

Infrared spectral observations are crucial to understanding the nature of obscured sources of luminosity in the universe.  For many galaxies, the great majority of the emitted luminosity arises in the infrared as emission from dust which has been heated by hot stars or an active galactic nucleus (AGN) which is heavily obscured at optical wavelengths. In such cases, the nature of the primary source can only be determined indirectly with clues derived from the infrared emission.  This determination is especially important for objects at high redshift (z $\ga$ 1), where it has been established that the bulk of luminosity density in the universe derives from such infrared dust emission \citep{hau98}, but it is not established whether this emission is powered primarily by starbursts or by AGN.  Understanding the origin and evolution of high redshift sources crucially depends, therefore, on interpretations of infrared characteristics, especially for the most luminous sources, or ULIRGS (e.g. \citet{far03}).  It is clear that starbursts and AGN are both important, but it is not yet clear how to make a definitive classification for individual objects.  With the new capabilities of the IRS on Spitzer, and the possibility of obtaining comprehensive spectra of large samples to high redshift, it is important to examine in more detail the spectral characteristics of known AGN in the accessible spectral range of $\sim$5\,\um to $\sim$37\,\um. 

Over the past three decades, much progress has been made in understanding the various observed manifestations of AGN at all wavelengths. Certain AGN, because of brightness, proximity, or early discovery, have received extensive attention.  The objects selected for analysis in the present paper are all very well studied and are chosen because they represent a wide range of AGN classifications derived from wavelengths other than infrared. The primary objective of comparing them is to determine to what extent differences in the infrared spectra can be related to other characteristics already known. It is especially crucial to identify any diagnostics based only on spectral morphology which can characterize AGN. The classification scheme for AGN which defined the type 1 and type 2 categories was based on a simple spectral classification which was possible because the optical portion of the spectrum shows both permitted and forbidden emission lines.  When the hydrogen permitted lines were broader than [OIII] forbidden lines, it was Seyfert 1; if [OIII] and hydrogen lines were the same width, it was Seyfert 2 \citep{kw74}.  More refined intermediate classifications, such as Seyfert 1.5 \citep{ost81}, related to the actual widths and relative strengths of permitted compared to forbidden lines.  While these were initially simple morphological classifications, they subsequently proved to have significant astrophysical meaning because the permitted lines can arise in much denser regions than the forbidden lines; this denser region only showed in the broad wings of the Balmer lines, thereby defining the "broad line region"(BLR).  The unifying interpretation of differences between Seyfert 1 and Seyfert 2 as arising only from orientation effects, with the BLR hidden in Seyfert 2 by an intervening torus, arose from the observation that the hidden BLR can be seen in Seyfert 2 from scattered, polarized light \citep{ma83}. 

In the present paper, we present full low resolution and high resolution spectra for a set of 8 classical but diverse AGN of various types: NGC 4151, NGC 7469, NGC 1275, NGC 3079, Markarian 3, Markarian 231, I Zwicky 1, and Centaurus A.  Various features are measured, and parameters related to the shape of the continuous spectrum and the strength of gaseous, molecular, and solid-state spectral features are defined. The dispersion among these parameters is examined for these AGN and compared to the same parameters seen in important starburst sources. The approach is also morphological, to examine various characteristics of the spectra to determine any correlations with AGN type or as discriminant between AGN and starbursts. Even though all of these sources have an extended galaxy component, we extract spectra only from the bright nucleus that is spatially unresolved by Spitzer to obtain a collection of spectra that are dominated by AGN characteristics with as little contamination as possible from the surrounding galaxy.  

Spectra for the mid-infrared have the disadvantage of containing no strong permitted emission lines, so there are no detectable lines arising in the BLR.   Because they are forbidden lines, the strongest observed infrared emission lines arise in the narrow line region, which means their strength depends in a complex way on how much ionizing radiation escapes the BLR and how much additional ionization arises from circumnuclear starbursts.  Nevertheless, the best mid-infrared AGN diagnostic derives from the presence of high-ionization emission lines from the narrow line region \citep{stu02}. We present high resolution observations of all sources to evaluate this diagnostic.  Primarily, however, we desire a scheme that is useful for classifying distant, faint sources when only poor S/N infrared spectra are available. It is especially crucial for these sources to determine whether the mid-infrared luminosity arises primarily from an AGN or a starburst, but lines are generally not detectable in faint sources for which only low resolution spectra of low S/N are available.  Correlations with properties of AGN at other wavelengths indicate that much of the mid-infrared continuum arises from dust heated by the AGN luminosity, so there is the possibility that characteristics of the dust continuum might be an AGN indicator. We examine the spectral morphology of our sample of AGN to define parameters that might aid in classification of other sources based only on the continuum shape or the strength of strong spectral features.  We emphasize parameters applicable to faint sources at high redshift (1 $<$ z $<$ 3) because such sources are already being discovered with the IRS (\citet{hou05}, \citet{yan05}, \citet{lut05}), so we especially examine parameters within the low resolution spectra at wavelengths $\la$ 15\,\um.  Although none of the AGN described here would be sufficiently luminous to be detectable with Spitzer at such redshifts, it is reasonable to anticipate that the infrared spectral characteristics of AGN would remain similar even at much greater luminosities, as proved to be the case when scaling optical spectra from Seyfert 1 galaxies to type 1 quasars. By providing a variety of AGN "templates", we can increase the ability to divide any given source into AGN and starburst components (e.g. \citet{tra01}, \citet{far03}). A companion paper \citep{bra05} presents a variety of starburst templates which we have observed with the IRS.

\section{Observations}

These eight AGN were among the first observed with the IRS\footnote{The IRS was a collaborative venture between Cornell
University and Ball Aerospace Corporation funded by NASA through the
Jet Propulsion Laboratory and the Ames Research Center} \citep{hou04}. All modules were used for all objects, which includes both orders of the low resolution short wavelength module (SL1, SL2), both orders of the low resolution long wavelength module (LL1, LL2), the high resolution short wavelength module (SH), and the high resolution long wavelength module (LH).  Details of the observations are presented in Table 1. The sum of all observations of all objects reported in this paper required a total of 1.6 hours integration time with the various modules of the IRS. The SSC pipeline which applies calibration to the observations has undergone numerous revisions since the observations were made.  The spectra presented here have been reprocessed since the original observations and are based on pipeline version S11 which applies dark corrections, droop and non-linearity corrections, and flat fields.  Spectra are extracted from the two-dimensional pipeline images using the SMART package \citep{hig04}, and final calibration to flux units is determined by scaling to the calibrating star $\alpha$ Lacertae for low resolution and $\xi$ Draconis for high resolution, extracted identically in SMART. Background subtraction for SL1, SL2, LL1 and LL2 modules was achieved by subtracting the background observed in the off-source order from the on-source observation; e.g., the LL1 slit provides an observation only of background when the source is in the LL2 slit, and this background is subtracted from the observation when the source is in
the LL1 slit.  Within the spectral extraction process, we can measure the width of the source in the direction perpendicular to dispersion, and all of the sources included here show an unresolved nucleus even though there may be detectable extended emission surrounding the nucleus. The spectral extractions are designed for unresolved point sources, so that the spectrum of the bright nucleus is extracted by itself even when there is extended emission surrounding it. The  spectra represent the nuclei within FWHM of $\sim$ 3.5'' (two pixels) at 10\,\um.  The widths of extracted spectra are diffraction limited so scale with wavelength.  The SL slit is also narrower (3.6'') than the LL slit (10.6''). The net result is that the "aperture" applied to an unresolved nucleus in LL at 20\,\um, for example, has six times the area of the "aperture" applied in SL at 10\,\um.  If a source is completely unresolved, this aperture difference should not cause any difference in flux levels at the wavelength where SL spectra are mated, or "stitched", to LL spectra.  Conversely, an extended source of constant surface brightness would appear brighter in LL in proportion to the difference in aperture area.  For the AGN spectra, occasional differences are found in the flux levels within SL spectra compared to LL.  To normalize spectra across the full wavelength range covered by SL and LL, all spectra presented here are scaled to the extracted flux of the LL1 spectrum. The scaling factors required to increase SL to match LL1 are 1.08 for NGC4151, 1.15 for Markarian 3, 1.00 for I Zw 1, 1.03 for NGC 1275, 1.11 for Centaurus A, 1.16 for NGC 7469, 1.05 for Markarian 231, and 1.49 for NGC 3079.  It is not surprising that the correction is greatest for NGC 3079, because this is a low luminosity nucleus, faint compared to the surrounding galaxy, so the relative contribution of extended, circumnuclear emission would be greater. That the stitching corrections are small for the other sources gives confidence in the assumption that the extracted spectrum primarily represents the unresolved AGN, so we conclude that any spatially resolved circumnuclear emission is not a significant source of contamination for the "nuclear" spectra presented, except for NGC 3079.

For the high resolution observations, we have no independent background observation to subtract background continuum from source continuum.  This means that we do not have a measure of the source-only continuum for the high resolution spectra as is measured for the low resolution spectra.  The shorter wavelength slit for high resolution (SH; 9.9\,\um to 19.6\,\um) is 4.7'' by 11.3'' whereas the longer wavelength slit (LH; 18.7\,\um to 37\,\um) is 11.1'' by 22.3''. Spectral extractions include all of the flux in the slits. This means that the level of the observed continuum can be greater in LH just because of increased background included in the slit at the longer wavelengths even if all of the source flux is from an unresolved nucleus, with no surrounding circumnuclear emission in the source.   As a result, we do not attempt to correct the high resolution spectra for differences in the continuum level and show these differences in the Figures. It can be seen from the Figures that the continuum levels of high resolution and low resolution are, in fact, very consistent, which implies that most of the continuum seen in high resolution  indeed arises from the unresolved AGN.  The source with the greatest discrepancy, and also the greatest difference between SH and LH, is I Zwicky 1.  This is the faintest source, so these discrepancies are consistent with having an increased relative contribution of background to the observed high resolution continuum.  Strong emission lines are detected in the low resolution spectra (Figures 1-8) so that reliable equivalent widths of the emission lines, based on the true source continuum, can be determined from the low resolution spectra.  The line to continuum ratio is parameterized by the equivalent width of [Ne III] 15.6\,\um as measured in low resolution, because this is an intermediate excitation line conspicuous in starbursts and AGN and which is not contaminated by nearby features in low resolution spectra. (For Markarian 231, this line is too weak to be seen in low resolution so the equivalent width is determined by comparing the line flux measured in high resolution with the continuum at that wavelength determined from low resolution.) 

Actual emission line fluxes are determined from the high resolution spectra because the underlying continuum can be more accurately subtracted, especially for weak lines.  The line fluxes are measured independently in each of the two high resolution modules (SH and LH) with no effort to normalize these using any scaling from the high resolution continuum.  Measured values for emission lines are in Table 3, although we have not included all lines that can sometimes be seen in the high resolution spectra.  We note, for example, that we do not include [FeII] 25.99\,\um measures or limits, although we have been careful to distinguish this feature from the adjacent [OIV] 25.89\,\um.  Relative line intensities are normalized to the flux of [Ne II] 12.8\,\um for which the actual fluxes and equivalent widths are listed in the Table footnote. Each observation produces two independent spectra, for two slightly different positions on the slit (the two "nod" positions).  To illustrate our measurement uncertainty, line measures from each nod are listed. If a line appears within two orders of the high resolution spectrum with comparable S/N, the average of both values is listed for each nod.  If the line is only within one order or has significantly better S/N in one order, only that order is used. In the footnotes to Table 3, comparisons are given with Infrared Space Observatory (ISO) measurements \citep{stu02} of the [Ne II] 12.8\,\um flux for the six objects in common. There are discrepancies of about a factor of two for Markarian 3, I Zwicky 1, and NGC 1275 for the absolute values of some line fluxes, although relative intensities are in better agreement.  We attribute these differences between Spitzer and ISO measurements primarily to the challenges of measuring weak lines in sources of poor signal to noise for which the definition of the continuum level is difficult. The limiting factor in the ability to detect weak lines is the 
contrast compared to the continuum rather than the actual flux of the line.  The limit depends, therefore, on the noise within the continuum.  We determine limits by extracting the largest "feature" in the continuum at the wavelength of the line and find that the flux limits listed correspond to an average equivalent width limit of 0.0015\,\um within the high resolution spectra; this is an empirical limit even though all of the continuum may not arise from the source itself. 

\section{Results}

Full spectra of the eight sources are shown in Figures 1 through 8 as observed with all modules and orders of the IRS . Considering the full spectra in Figures 1-8, the most conspicuous differences are in the equivalent widths of the emission lines and in the slopes of the continuum for wavelengths $>$ 20\,\um.  We define spectral parameters useful with low S/N spectra, such as from sources at high redshift, for which it is not always possible to determine specifically if the continuum is affected by any broad absorption or emission features.  Therefore, we desire some parameters that require only a measure of the flux density at specific rest wavelengths without judging the presence or absence of spectral features. This is a complex portion of the spectrum. Dramatic differences are evident in the strength of the silicate feature, ranging from strong emission to deep absorption. Also evident are differences in the overall slope of the continua and in the strength of the PAH emission. The broad and continuous spectral coverage available with the IRS makes it possible to determine continuum flux measures at wavelengths relatively unaffected by the various strong features.   The slope of the longer wavelength portion of the spectra is parameterized as a flux ratio in the continuum, using $f_{\nu}$(30\,\um)$/f_{\nu}$(20\,\um).   From examining the spectra illustrated, the "cleanest" location is at 15\,\um, between the two silicate features. Therefore, we define a parameter representing the slope of the underlying continuum using flux density measurements at 6\,\um and 15\,\um. The effect of any silicate feature, absorption or emission, is strongest at about 10\,\um. Flux densities at this wavelength thereby give a measure of the influence of any silicate features.  In summary, the parameters we introduce for describing the continuum are: 1. The slope of the longer wavelength portion of the spectra parameterized as a flux ratio using $f_{\nu}$(30\,\um)$/f_{\nu}$(20\,\um); 2. the underlying continuum shape shortward of 15\,\um parameterized by $f_{\nu}$(15\,\um)$/f_{\nu}$(6\,\um); 3. the continuum including the effect of any silicate feature parameterized by $f_{\nu}$(10\,\um)$/f_{\nu}$(6\,\um). These continuum parameters are listed in Table 2 and fluxes of the most significant emission features are listed in Table 3. Objects are ordered in the Figures and Tables by the continuum slope as reflected in the $f_{\nu}$(30\,\um)$/f_{\nu}$(20\,\um) parameter, with the flattest spectra described first.

Brief summaries of the most important observed characteristics of the individual sources follow:

NGC 4151 is a classical Seyfert galaxy and is the optically-brightest example of an AGN with a visible broad line region.  The spectrum in Figure 1 shows a flattening of the continuum at about 20\,\um, implying a conspicuous hot dust component. There are weak excesses above a smooth continuum probably caused by both 10\,\um and 18\,\um silicate emission features.  The high ionization lines of [NeV] and [OIV] are very strong, indicating that the narrow line region is primarily ionized by the AGN (\citet{stu99}, \citet{ale99}). There is no evidence of PAH emission features. The periodic "fringing" in the continuum longward of $\sim$ 20\,\um in this and other spectra is an instrumental artifact caused by interference in the LL1 interference filter and in the substrate of the long low detector. This is not removed during calibration if an unresolved source is located slightly differently in the slit from the calibration source used to determine the overall spectral response. Algorithms are available to remove this fringing, but we did not attempt to do so in order to avoid the possible introduction of any spurious features, or the removal of any real features. 

Markarian 3 is included to represent a Seyfert 2 AGN, without a visible broad line region, because the prototype Seyfert 2, NGC 1068, is too bright for a full spectrum with the IRS. The spectrum in Figure 2 shows a remarkable similarity to NGC 4151 in terms of eqivalent width of emission lines and the flattening near 20\,\um. [NeV] is also strong, indicating ionization of the narrow line region by the AGN. There is an important difference, however, in that Markarian 3 shows a depression of the continuum near 10\,\um attributable to silicate absorption, whereas NGC 4151 has no silicate absorption but probably has silicate emission.  

I Zwicky 1 is sometimes called a QSO (PG0050+124) and sometimes a "narrow line" Seyfert 1 galaxy and is a luminous but radio quiet AGN.  The IRS spectrum in Figure 3 shows strong silicate emission already discussed \citep{hao05}. Emission line equivalent widths are weak, and there are no PAH features. The spectrum "break" at 20\,\um is not present. 

NGC 1275 (Perseus A) is a strong and variable radio galaxy with an optically visible AGN which can be considered a Seyfert 1 based on the presence of broad, although weak, permitted lines \citep{fs85}.  The spectrum in Figure 4 clearly shows 10\,\um silicate emission and probably 18\,\um emission.  Emission lines are present but weak, and there are no PAH features.  Because of the compact and variable radio emission, this source might be expected to have significant non-thermal continuum, but the presence of the silicate feature indicates unambiguously that much of the mid-infrared continuum arises from dust emission. 

Centaurus A is the closest AGN and a luminous radio galaxy but with a nucleus completely obscured optically; a mid-infrared ISO spectrum of the nucleus is in \citet{mir99}.  This source is resolved along the slit of the IRS, but the spectrum shown is only that of the central nucleus within the same IRS spatial resolution as an unresolved source. Strong silicate absorption is present in Figure 5, and the emission lines are also strong.  The strength of [Ne V] indicates ionization by the AGN, but PAH features are also present.  Particularly notable is the strength of the complex of features near 17\,\um similar to the complex of molecular Hydrogen and PAH features seen in NGC 7331 \citep{smi04} and discussed by \citet{pee04a}.  These are much stronger relative to the shorter wavelength PAH features than in other objects with PAH features which are discussed herein.  

NGC 7469 was one of the original Seyfert galaxies \citep{sey43}, subsequently classified as Seyfert 1, and later found to have a strong circumnuclear starburst \citep{wil86}.  The IRS spectrum in Figure 6 shows a continuum similar to that of the pure starburst NGC 7714, including strong PAH features.  However, there is a weak [Ne V] line indicative of some contribution to ionization from the AGN. There is no evidence of silicate absorption or emission.  This is a clear example of how a starburst signature can dominate the infrared spectrum even if an AGN dominates the optical classification. 

Markarian 231 is a Seyfert 1 because of the broad Balmer lines, but it is the AGN with the strongest optical signature of absorbing material, having very deep, blueshifted sodium absorption \citep{aw72}, and it was also one of the first ultraluminous infrared galaxies discovered \citep{rl72}.  It is the low redshift prototype for a "broad absorption line" AGN, with extensive X-ray studies \citep{gal02}. It has a luminous circumnuclear starburst considered responsible for most of the ULIRG luminosity \citep{sol92}. The IRS spectrum in Figure 7 shows only weak PAH features and is dominated by very strong silicate absorption; there are no detectable emission lines in low resolution, although the fringing may obscure some lines. The long wavelength spectrum is steeper than for any of the AGN previously discussed, perhaps attributable to cooler dust associated with the starburst. An unusual feature in Markarian 231 is the broad excess of emission centered at $\sim$ 8\,\um, blueward of the silicate absorption feature.  This excess is common in luminous, high redshift sources for which Markarian 231 provides a good template fit \citep{hou05} so the feature is important to understand.  It does not show the structure of the PAH features at similar wavelengths in NGC 7469 and NGC 3079, and the emission excess is much greater compared to the weak 6.2\,\um PAH feature than for the relative flux of the 7.7\,\um PAH feature in the starbursts. All of the other starburst galaxies we have observed show PAH features similar in relative flux and structure to those in NGC 3079 \citep{bra05}. A similar excess blueward of the absorption in Centaurus A can be attributed to PAH emission because it has the structure and correct relative flux compared to the 6.2\,\um feature. 

NGC 3079 is a Seyfert 2, but the AGN classification is not conspicuous in the optical and was found only through a targeted search of nearby galaxies for low luminosity AGN \citep{ho97}.  This is an edge-on galaxy having a nuclear starburst with a supernovae-driven wind emerging out of the disk \citep{vei94}. This has by far the strongest PAH features of any object in the current sample and also has strong low-ionization emission lines.  Because of the strong PAH features, the underlying continuum fit, free from any PAH contribution, is very uncertain shortward of 18\,\um (panel A in Figure 8).  A continuum could be fit to wavelengths from $\sim$ 5\,\um to $\sim$ 20\,\um without requiring silicate absorption near 10\,\um. This source indicates the importance of having a long continuum baseline for sources with PAH features.   If the spectrum covered only the region $<$ 15\,\um as in Figure 8C, the spectrum would indicate strong absorption at 10\,\um.

\section{Interpretation}

Our primary motive for obtaining these spectra is to have a set of mid-infrared comparison spectra for well known AGN whose character is established from observations at other wavelengths.   We especially desire to determine if there are characteristics of the mid-infrared spectrum which are unique to AGN and which can determine unambiguously whether the primary luminosity source responsible for heating the dust is an AGN or a starburst.  This question has been addressed with extensive samples from ISO observations (\citet{gen98}, \citet{stu02}, \citet{lau00}) using as primary diagnostics the strengths of high excitation emission lines, the slope of the continuum, and the strength of the PAH features. These ISO results indicate some trends:  sources with AGN have relatively stronger high excitation lines, weaker PAH features, and flatter spectra (implying hotter dust) for wavelengths $<$ 10\,\um.  While useful, these trends do not always lead to an unambiguous classification, as can be seen from examples in the current study.  

Perhaps the most fundamental conclusion from comparing these 8 objects is that each spectrum is unique, which is not surprising given the deliberate choice of objects having a variety of AGN classifications.  As examples, Markarian 3 and NGC 4151 look similar in terms of continuum slopes and emission line characteristics, but Markarian 3 shows a silicate absorption and NGC 4151 weak silicate emission. Centaurus A and I Zwicky 1 show similar spectral slopes $<$ 15\,\um, but Centaurus A has deep silicate absorption with PAH features and strong emission lines, whereas I Zwicky 1 has strong silicate emission and no PAH or emission lines.  NGC 7469 shows strong PAH features indicative of starburst dominance, but it has a flatter continuum $<$ 15\,\um than the AGN of NGC 1275 and Markarian 3.  The many differences among the spectra arise because of the many physical parameters that control the mid-infrared spectra.  These include the mix of dust temperatures and emissivities, composition of the dust, optical depths of the different dust components, optical depths to the emission lines and other features, and ionizing source of the emission lines.  Each object can, in principle, be modeled to determine these parameters as uniquely as feasible. We do not present such modeling in the present paper, because our objective here is only to present the spectra and seek overall correlations that might aid in classifying much larger samples of objects with infrared spectra for which comprehensive data at other wavelengths is not available. 

Another fundamental result is that there are no featureless spectra in the mid-infrared; there are no smooth power laws.  Those objects which do not have strong emission lines or PAH features show silicate features, either in emission or absorption.  In the spectral region from 6\,\um to 15\,\um, a power law can be fit to a continuum baseline on which the features are superposed.  This power law is well represented simply by the flux densities measured at 6\,\um and 15\,\um, where there are no strong features. (The exception is NGC 3079, for which the PAH features are so strong that location of the underlying continuum is very uncertain.) The corresponding spectral index of the power law is given in Table 2 and has mean and dispersion of 1.77 $\pm$ 0.60.  

The observed spectra are generally consistent with the "unified scheme" for AGN, whereby the difference between type 1 and type 2 is assigned to the orientation effects of an obscuring torus.  For the Seyfert 1 AGN NGC 4151, I Zwicky 1 and NGC 1275, the silicate is in emission, but it is in absorption for the Seyfert 2 Markarian 3, as expected \citep{pk93}.  The "broad absorption line" type 1 AGN Markarian 231 shows strong silicate absorption and no emission, but the broad line region of this unusual source is heavily obscured although optically detectable \citep{gal05}.  We noted the curious emission hump blueward of the silicate absorption.  This apparent emission begins to rise at $\sim$ 7.5\,\um, which is where the silicate opacity spectra of \citet{wei01} and \citet{li01} predict silicate emission should first appear, and which is consistent with the observed beginning of the silicate emission profile in the two type 1 AGN with the strongest silicate emission (PG 1351 and I Zwicky 1 in \citet{hao05}). To suppress the much stronger emission that should then appear at longer wavelengths, the silicate emission profile would have to be strongly self absorbed.  An alternative interpretation is that applied to the deeply obscured sources NGC 4418 \citep{pee04b} and IRAS F00183-7111 \citep{spo04} for which the apparent hump of emission is actually the underlying continuum which has been absorbed by ice and hydrocarbon features shortward of 7.5\,\um.  Either interpretation attributes this unusual feature to effects of absorption rather than PAH emission.

We have a large sample of objects defined from optical classification as starbursts whose IRS spectra are presented elsewhere \citep{bra05}. Using the various parameters introduced, we can seek any that would clearly separate starbursts from AGN, especially for which we can observe the rest frame continuum to z $\sim$ 2.   We also compare with IRS spectra for SBS 0335-052 \citep{hou04b}, which is a very metal poor, compact starburst in a dwarf galaxy. Because of the low metallicity and its compact nature, it may be the best available representative of a primordial starburst.  (The parameters for SBS 0335-052 have been measured from improved versions of the spectra compared to those previously published, processed with the S11 pipeline, and are listed in Table 2.) The comparison used is similar to the diagram comparing continuum slope and PAH equivalent width of \citet{lau00}. The parameter $f_{\nu}$(15\,\um)$/f_{\nu}$(6\,\um) represents the continuum independent of any strong features and should be a measure of the influence on the continuum of the hottest dust component, with sources having hotter dust showing flatter spectra, so the slope of this continuum should be affected by the presence of an AGN.  PAH emission is interpreted as arising from photodissociation regions within star-forming clouds so should be an indicator of starbursts.  We show in Figure 9 the comparison of these parameters for the 8 objects in the present paper and the 21 objects in \citet{bra05}. This diagram confirms that the strength of the PAH features measured as equivalent width is a clear discriminate between AGN and starbursts, extending even to intermediate cases.  This conclusion must include an important caveat, however, which is that the compact, metal poor starburst of SBS 0335-052 cannot be distinguished from an AGN in this diagram because of its very weak PAH features \citep{thu99}. The general result from \citet{lau00} indicated that the ratio of "warm" (14\,\um- 15\,\um) to "hot" continuum (5.1\,\um- 6.8\,\um) is also a discriminate between starbursts and AGN.  However, the fact that the $f_{\nu}$(15\,\um)$/f_{\nu}$(6\,\um) parameter in Figure 9 shows a similar range for AGN and starbursts indicates that a flatter continuum is not an infallible criterion for separating AGN and starbursts.  For use in other comparisons to various samples of objects, we also measured continuum parameters $f_{\nu}$(30\,\um) and $f_{\nu}$(20\,\um).  The nature of the continuum described with these parameters is compared for AGN and starbursts in \citet{bra05}.

The presence of AGN is best verified by the detection of high ionization lines such as [NeV] and/or [OIV] \citep{stu02}.  Many of the most interesting infrared sources are not sufficiently bright to allow the necessary high resolution spectra for detecting these lines, and can also be at sufficiently high redshifts that useful diagnostic lines are redshifted out of the IRS spectral window.  It is desirable to determine from nearby AGN, therefore, if the high-ionization diagnostic generally relates to the PAH diagnostic discussed above. Within the present sample of 8 AGN, the ratio [OIV] 25.9\,\um/ [Ne II] 12.8\,\um in Table 3 ranges from 0.1 (NGC 3079) to 2.1 (Markarian 3). The lowest value would correspond to $<$ 10 \% AGN ionization and the highest to 100 \% in the mixing model of Sturm et al. This ratio is $<$ 0.09 in the prototype starburst NGC 7714, for example, consistent with its interpretation as a pure starburst \citep{bra04}.  All of the objects which would be classified as AGN in Figure 9 because of the small PAH equivalent width have [OIV] strengths or limits consistent with some contribution to ionization by an AGN.  All emission lines are so weak in I Zwicky 1 and Markarian 231 that [OIV] is not detected, even at a limit which would correspond to dominant AGN ionization.  The weakest detection in a "pure" AGN is for NGC 1275, with the ratio of 0.17 compared to [NeII] corresponding to $<$ 20 \% AGN ionization.  It is particularly encouraging that NGC 7469, which shows an intermediate starburst indicator in the PAH strength, would also indicate the presence of an AGN through the strength of [OIV], whose strength indicates $\sim$ 30 \% AGN ionization. While these trends need further statistical evaluation with larger samples of starbursts and AGN, these initial results with the IRS confirm that the degree of ionization in the narrow line region relates to the more easily measured PAH strength. 

\section{Summary}

Primarily as a comparison set for future observations of large samples of sources containing AGN or starbursts, low resolution and high resolution spectra for all wavelengths between 5\,$\mu$m and 37\,$\mu$m are presented for the well-studied AGN NGC 4151, Markarian 3, I Zwicky 1, NGC 1275, Centaurus A, NGC 7469, Markarian 231, and NGC 3079.  These spectra are characterised primarily by their diversity, and all differ significantly from others in some important features.  Some show strong silicate absorption (Markarian 231 and Centaurus A), some show strong silicate emission (I Zwicky 1 and NGC 1275) , some show strong PAH emission (NGC 7469 and NGC 3079), some show very strong emission lines (NGC 4151 and Markarian 3) and others very weak lines (I Zwicky 1 and Markarian 231).  Continuum slopes sometimes show significant breaks near 20 \um and other times do not; slopes of continua $<$ 15 \um within AGN are sometimes flatter (implying hotter dust) than for starbursts and other times not.  By comparison to a large sample of starburst galaxies, the results confirm previous conclusions (\citet{lau00}, \citet{gen98}) that the most consistent low-resolution spectral diagnostic for distinguishing an AGN-powered source from a starburst is the strength of the PAH features.  Much larger samples of IRS spectra of various sources already exist, so statistical refinements of these diagnostics should occur soon. 

\acknowledgments

We thank many others on the IRS team for
contributing to this effort, especially including G. Sloan, P. Hall, D. Barry, J. Bernard-Salas, J. Marshall, H. Spoon, L. Armus, and J. Higdon.  We also thank the referee, E. Sturm, for helpful suggestions.
This work is based on observations made with the
Spitzer Space Telescope, which is operated by the Jet Propulsion
Laboratory, California Institute of Technology under NASA contract
1407. Support for this work at Cornell University was provided by NASA through Contract
Number 1257184 issued by JPL/Caltech.

\clearpage

\begin{figure*}
\begin{center}
\resizebox{\hsize}{!}{\includegraphics{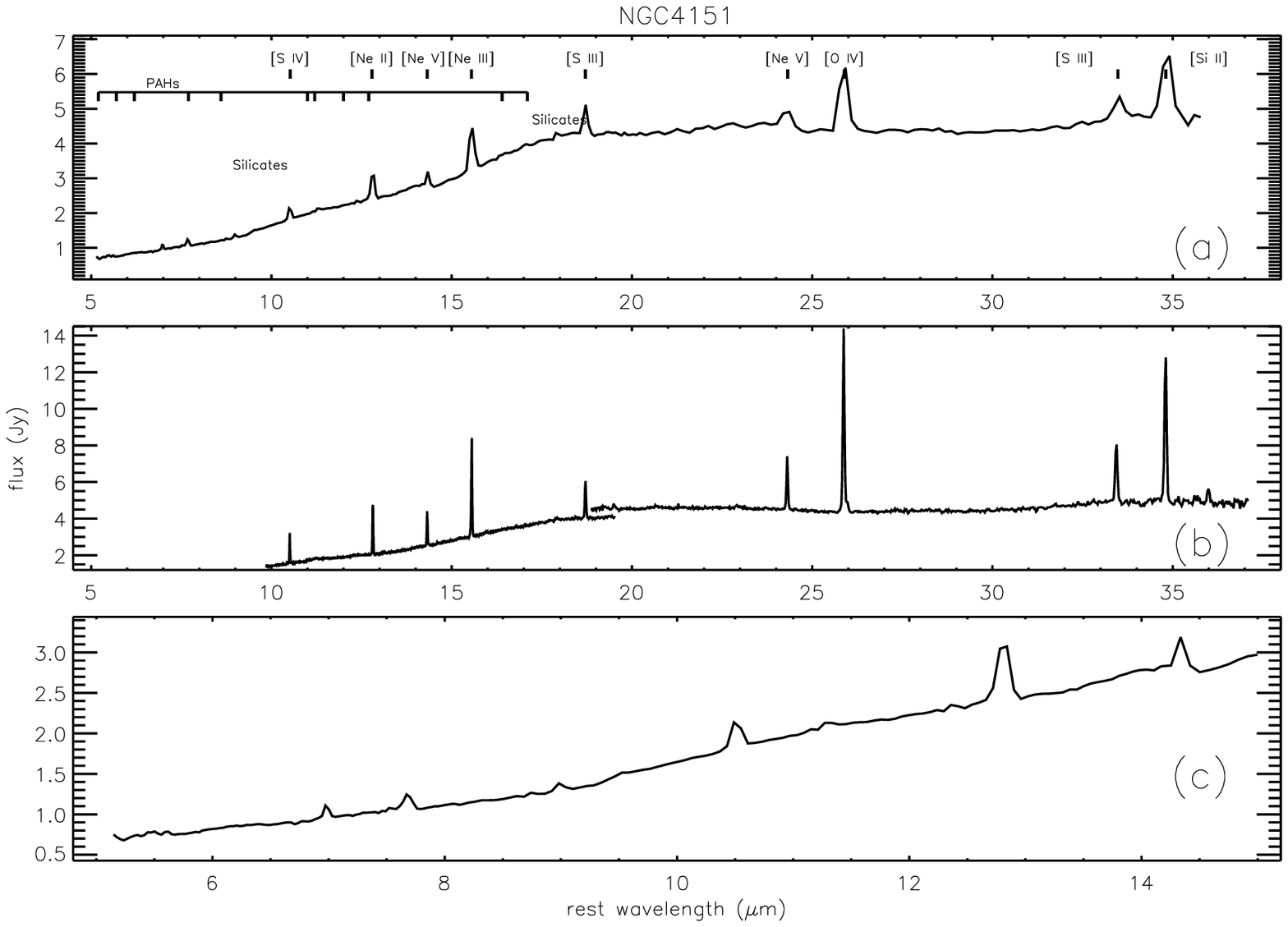}}
\caption{NGC 4151; panel A is full low resolution spectrum; panel B is full high resolution spectrum; panel C is enlarged 5\,\um to 15\,\um low resolution spectrum. 
\label{fig1}}
\end{center}
\end{figure*}

\clearpage

\begin{figure*}
\begin{center}
\resizebox{\hsize}{!}{\includegraphics{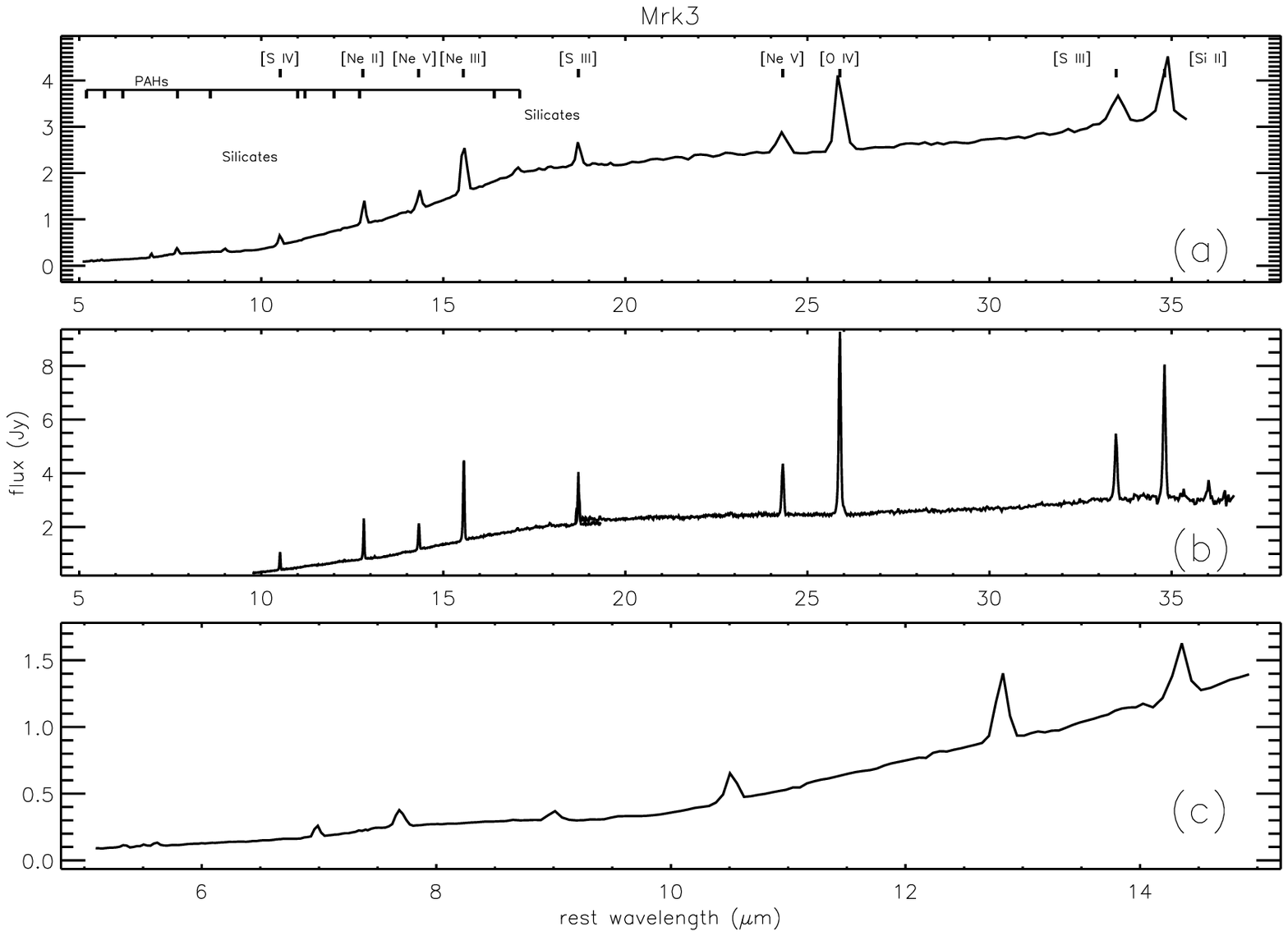}}
\caption{Markarian 3; panel A is full low resolution spectrum; panel B is full high resolution spectrum; panel C is enlarged 5\,\um to 15\,\um low resolution spectrum. 
\label{fig2}}
\end{center}
\end{figure*}

\clearpage

\begin{figure*}
\begin{center}
\resizebox{\hsize}{!}{\includegraphics{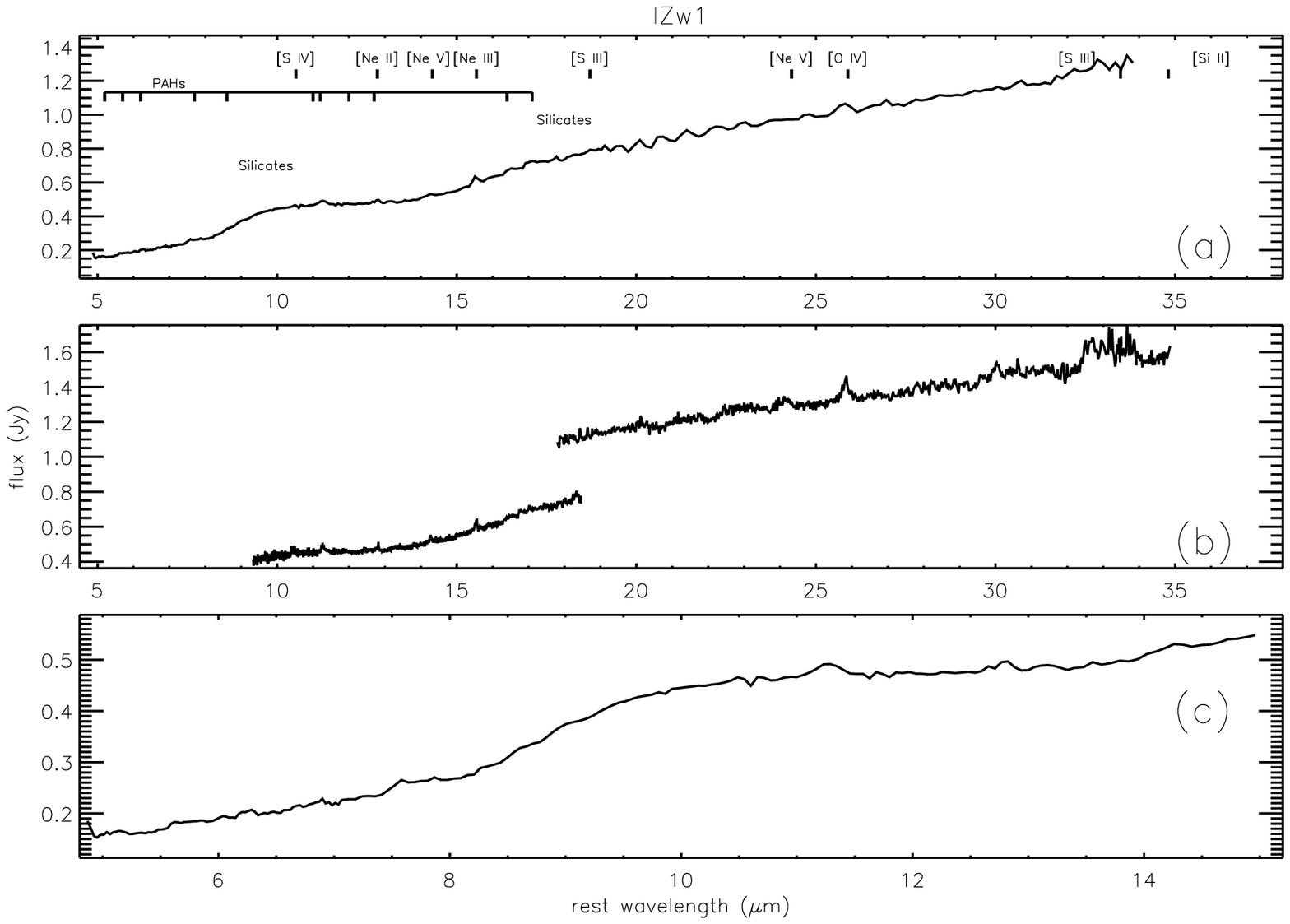}}
\caption{I Zw 1; panel A is full low resolution spectrum; panel B is full high resolution spectrum; panel C is enlarged 5\,\um to 15\,\um low resolution spectrum. 
\label{fig3}}
\end{center}
\end{figure*}

\clearpage

\begin{figure*}
\begin{center}
\resizebox{\hsize}{!}{\includegraphics{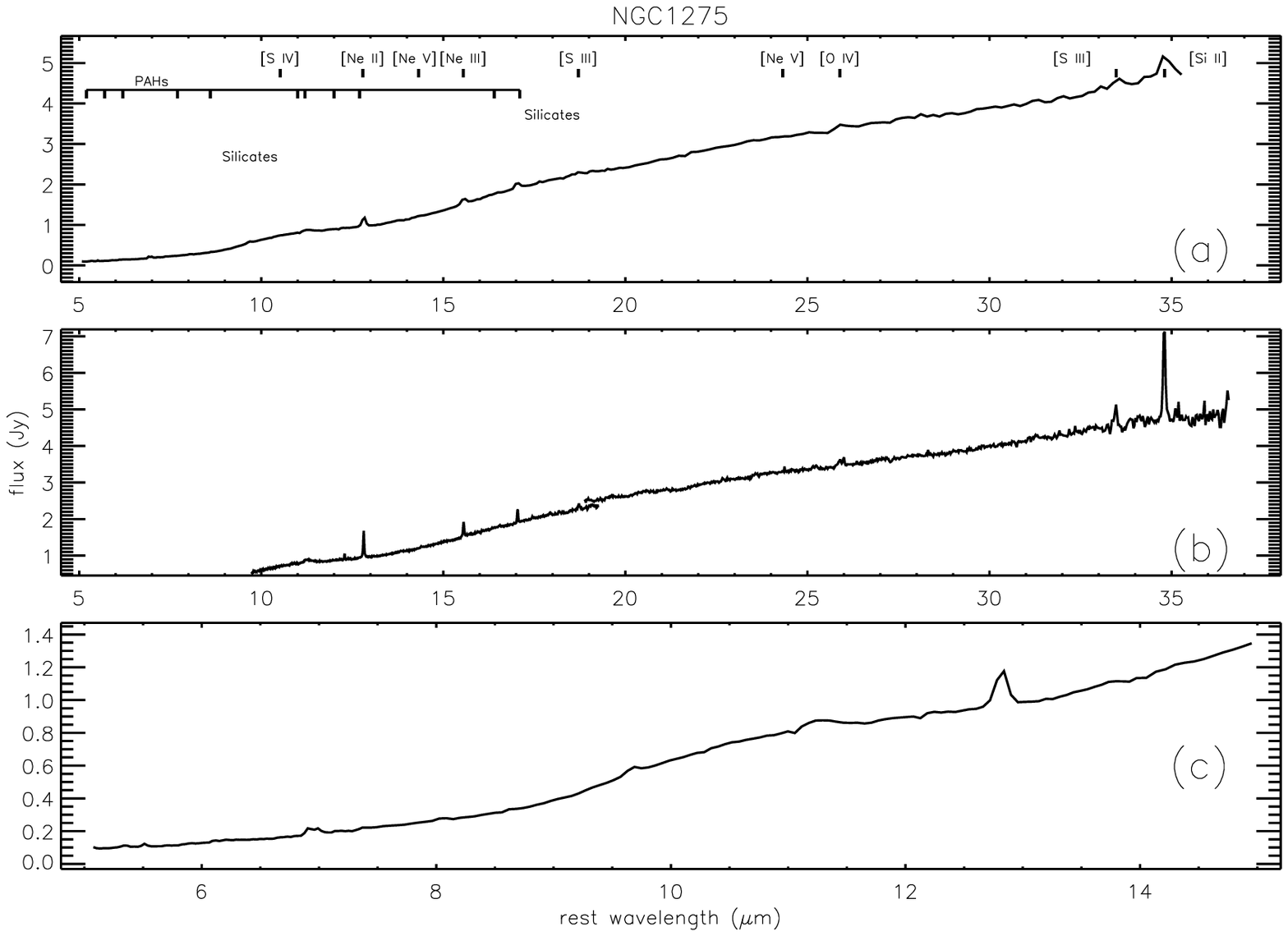}}
\caption{NGC 1275; panel A is full low resolution spectrum; panel B is full high resolution spectrum; panel C is enlarged 5\,\um to 15\,\um low resolution spectrum. 
\label{fig4}}
\end{center}
\end{figure*}

\clearpage

\begin{figure*}
\begin{center}
\resizebox{\hsize}{!}{\includegraphics{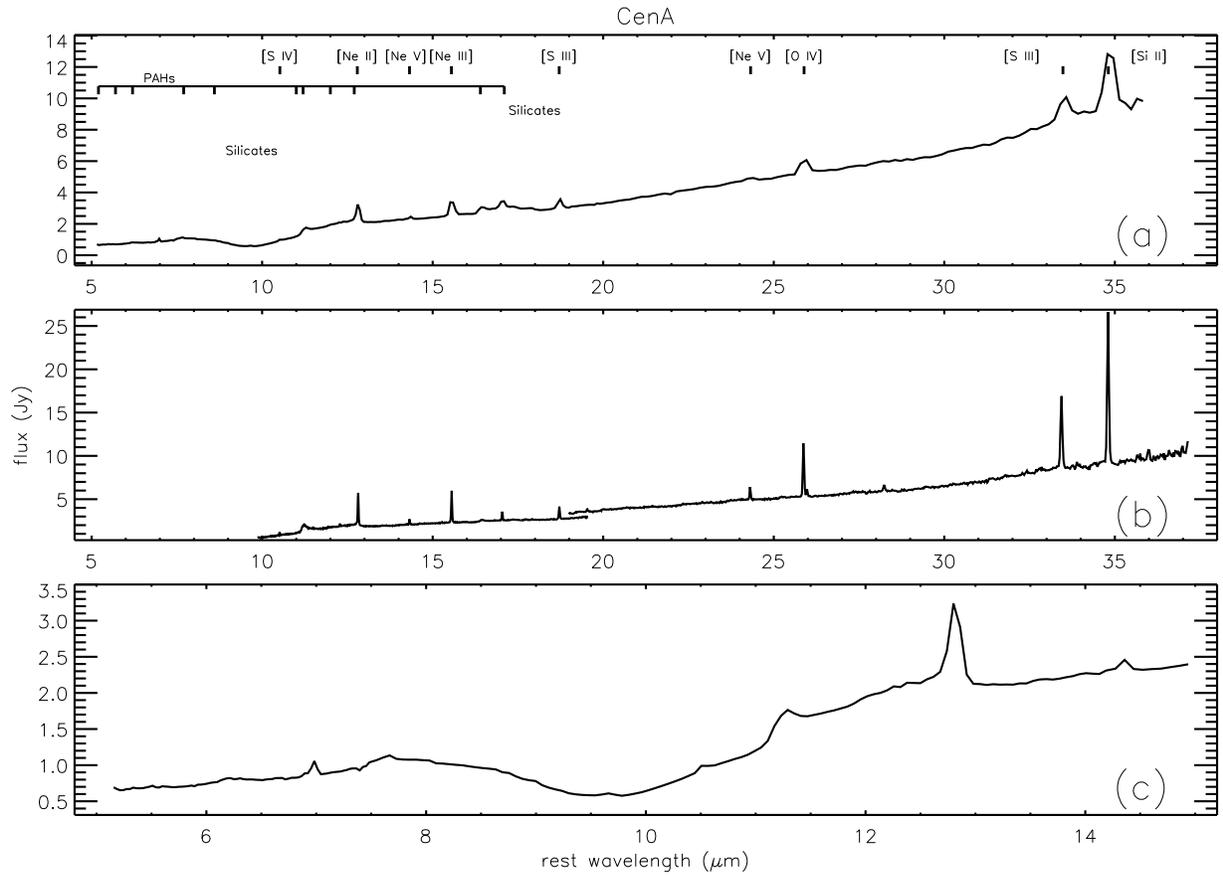}}
\caption{Centaurus A; panel A is full low resolution spectrum; panel B is full high resolution spectrum; panel C is enlarged 5\,\um to 15\,\um low resolution spectrum. 
\label{fig5}}
\end{center}
\end{figure*}

\clearpage

\begin{figure*}
\begin{center}
\resizebox{\hsize}{!}{\includegraphics{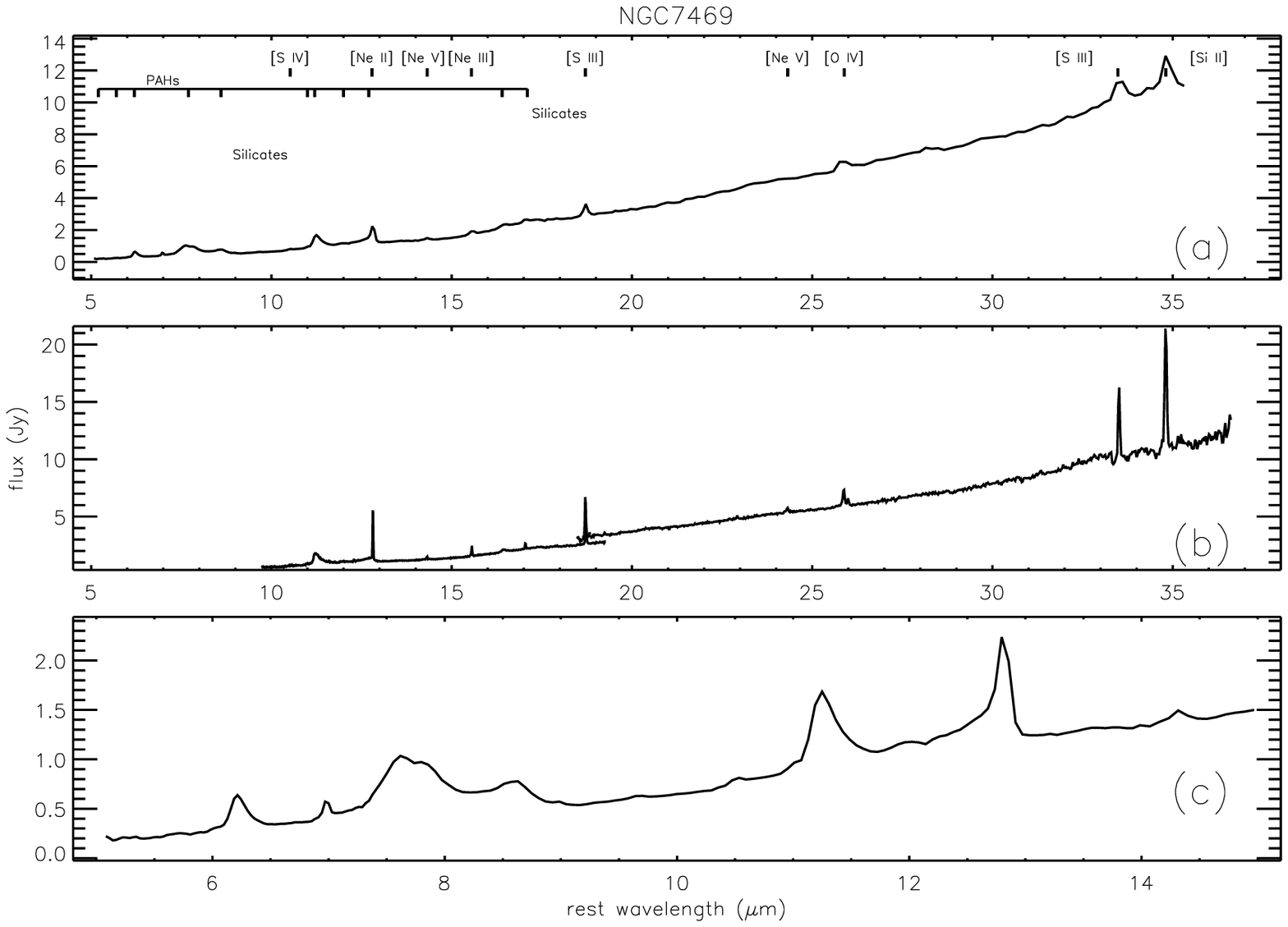}}
\caption{NGC 7469;  panel A is full low resolution spectrum; panel B is full high resolution spectrum; panel C is enlarged 5\,\um to 15\,\um low resolution spectrum. 
\label{fig6}}
\end{center}
\end{figure*}

\clearpage

\begin{figure*}
\begin{center}
\resizebox{\hsize}{!}{\includegraphics{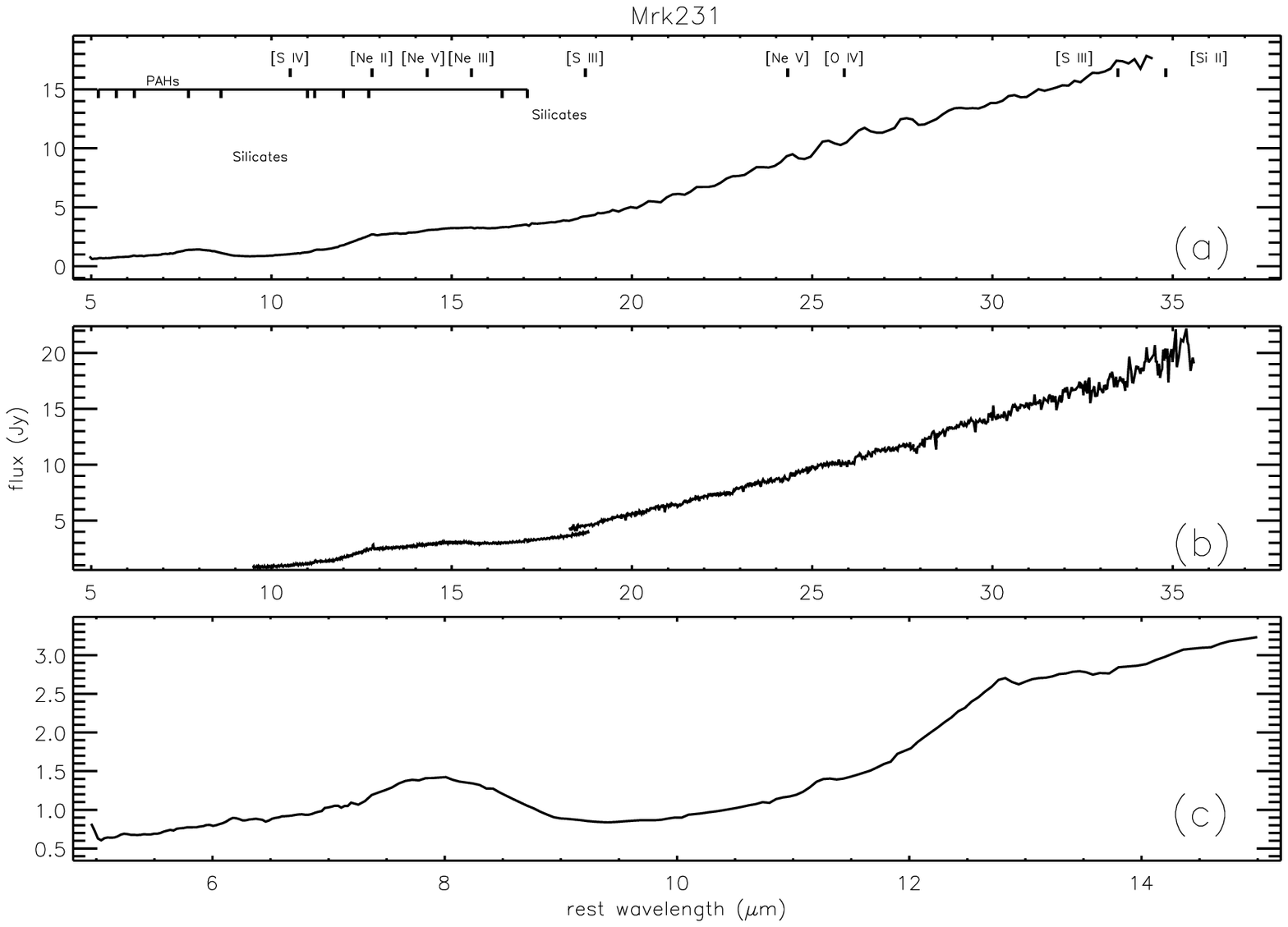}}
\caption{Markarian 231;  panel A is full low resolution spectrum; panel B is full high resolution spectrum; panel C is enlarged 5\,\um to 15\,\um low resolution spectrum. 
\label{fig7}}
\end{center}
\end{figure*}

\clearpage

\begin{figure*}
\begin{center}
\resizebox{\hsize}{!}{\includegraphics{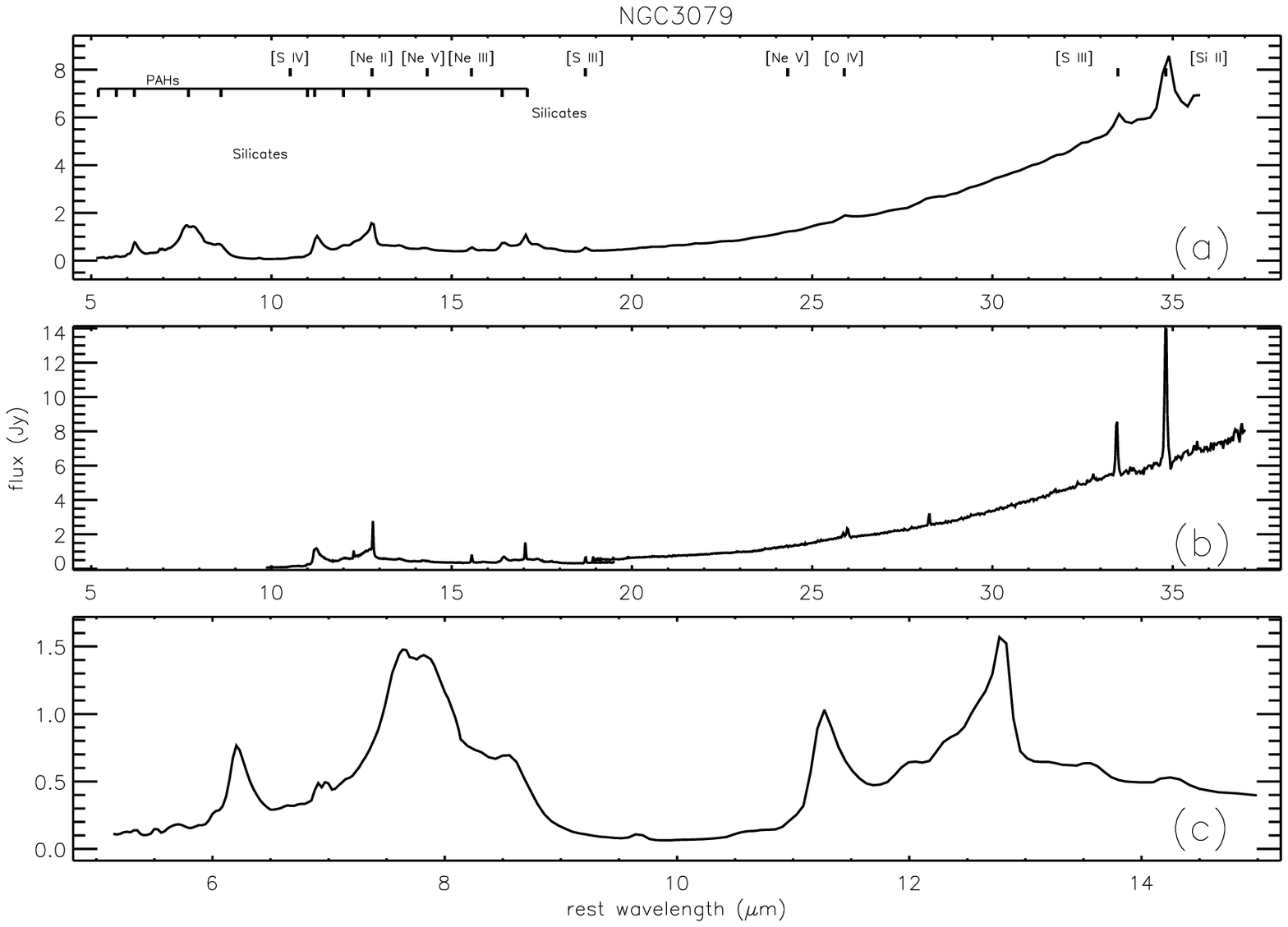}}
\caption{NGC 3079;  panel A is full low resolution spectrum; panel B is full high resolution spectrum; panel C is enlarged 5\,\um to 15\,\um low resolution spectrum. 
\label{fig8}}
\end{center}
\end{figure*}

\clearpage

\begin{figure*}
\begin{center}
\resizebox{\hsize}{!}{\includegraphics{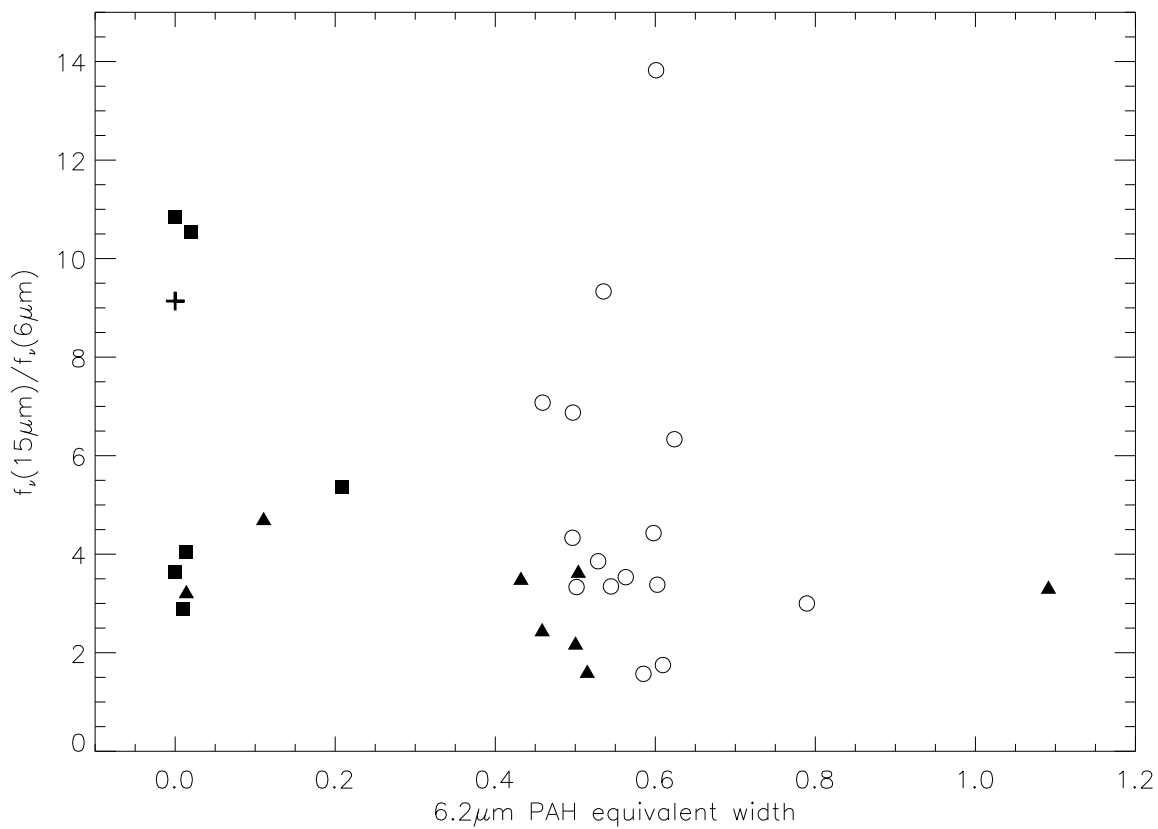}}
\caption{Comparison of continuum slope and equivalent width of PAH for AGN and starburst nuclei.  Filled squares: objects with conspicuous AGN indicators in optical classification which include NGC4151, Markarian 3, I Zwicky 1, NGC 1275, NGC 7469 and Markarian 231; triangles: objects with weak optical indicators of an AGN or AGN indicators from X-ray or radio characteristics which include Centaurus A, NGC 3079, Markarian 266, NGC 660, NGC 1097, NGC 1365,NGC 3628, and NGC 4945;  Open circles: sources with purely starburst optical classification which include NGC 7714, IC 243, Markarian 52, NGC 520, NGC 1222, NGC 2146, NGC 2623, NGC 3256, NGC 3310, NGC 3556, NGC 4088, NGC 4194, NGC 4676, NGC 4818, and NGC 7252; Cross: the compact, metal-poor starburst SBS 0335-052.
\label{fig9}}
\end{center}
\end{figure*}

\clearpage

\begin{deluxetable}{lccccccc}
\tablecolumns{8}
\tablewidth{0pc}
\tablecaption{Observations}
\tablehead{
\colhead{} & \colhead{} & \multicolumn{6}{c}{Integration times(sec)}\\
\cline{3-8}
\colhead{object} & \colhead{date} & \colhead{SL1} & \colhead{SL2} & \colhead{LL1} & \colhead{LL2} & \colhead{SH} & \colhead{LH}}
\startdata
NGC 4151    &  01/08/04 & 56 &    56 &              48 &    48  &           240  &          240 \\
Markarian 3 &  03/04/04 & 56 &    56 &            48 &    56 &            240  &          240 \\
I Zw 1      &  01/07/04 & 56 &    56 &              56 &    56  &           240  &          240 \\
NGC 1275    &  08/30/04 & 56 &    56 &              48 &    48  &           240  &          240 \\
Centaurus A\tablenotemark{1}& 02/27/04 & 48 &  48 &            48 &    48 &            240  &          224 \\
NGC 7469    &  12/16/03 &   56 &       56 &            60 &    60 &            240  &          240 \\
Markarian 231& 04/14/04 & 56 & 56 & 60 & 60 & 360 & 480\\
NGC 3079     & 04/19/04 & 56 &    56 &              56 &    56  &           240  &          240 \\
\enddata
\tablenotetext{1}{The Long Low observations of Centaurus A were on 07/14/04.}
\label{tab1}
\end{deluxetable}

\clearpage

\begin{deluxetable}{lcccccccccc}
\tablecolumns{11}
\tablewidth{0pc}
\tabletypesize{\tiny}
\setlength{\tabcolsep}{0.005in}
\tablecaption{Continuum Measurements}
\tablehead{
\colhead{} & \multicolumn{5}{c}{Continuum flux densities [Jy]} & \multicolumn{5}{c}{Continuum Ratios\tablenotemark{1}}\\
\cline{2-6} \cline{7-11} 
\colhead{object} & \colhead{$f_{\nu}$(30 \ums))} & \colhead{$f_{\nu}$(20 \ums)} &\colhead{$f_{\nu}$(15 \ums)} & \colhead{$f_{\nu}$(10 \ums)} & \colhead{$f_{\nu}$(6 \ums)} &\colhead{$f_{\nu}$(30 \ums)$/f_{\nu}$(20 \ums)} & \colhead{$f_{\nu}$(15 \ums)$/f_{\nu}$(6 \ums)} &\colhead{$\alpha$} &\colhead{$f_{\nu}$(10 \ums)$/f_{\nu}$(6 \ums)}& \colhead{$f_{\nu}$(30 \ums)$/f_{\nu}$(6 \ums)} }

\startdata
NGC 4151&  4.33&	 4.24&	2.98&	1.64&  0.82&  		1.02&    3.64&  1.40&  2.01&   5.3\\
Markarian 3&  2.72&	 2.19&	1.41&	0.36&  .13&  		1.24&    11.2&  2.62&  2.82&   21.5\\
I Zw 1	&  1.16&	 0.83&	0.55&	0.45&  .19&  		1.40&    2.89&  1.15&  2.34&   6.1\\
NGC 1275&  3.82&	 2.41&	1.36& 	0.64&  .129& 		1.59&    10.6&  2.56&  4.96&   29.7\\
Centaurus A&   6.4&	 3.3&	2.4&	 .64&  .74&  		1.93&    3.28&  1.29&  0.87&   8.7\\
NGC 7469&  7.8&	 3.2&	1.5&	 .65&  .28&  			2.39&    5.42&  1.83&  2.33&  27.9\\
Markarian 231& 14.4&	 5.3&	3.4&	 .95&  .84& 		 2.74&    4.06&  1.52&  1.13&  17.1\\
NGC 3079&  3.40&	 .48&	 .39&	 .06&  .24&  		7.13&    1.63&   \nodata  &  0.25&  14.3\\
SBS 0335-052& .084& .071&	 .053&	.0168&.0058& 		1.18&    9.1&   2.40&	2.9&  14.5\\
\enddata
\tablenotetext{1}{$\alpha$ is the spectral index for a power law that would connect $f_{\nu}$(15 \ums) with $f_{\nu}$(6 \ums).}
\label{tab2}
\end{deluxetable}
\clearpage

\clearpage

\begin{deluxetable}{lcccccccccc}
\tablecolumns{11}
\tablewidth{0pc}
\tabletypesize{\tiny}
\setlength{\tabcolsep}{0.005in}
\tablecaption{Strengths of Spectral Features\tablenotemark{1}}
\tablehead{
\colhead{} & \multicolumn{8}{c}{Relative Emission Line Strengths} & \multicolumn{2}{c}{Equivalent widths [\um]}\\
\cline{2-9} \cline{10-11} 
\colhead{Object} & \colhead{[SIV] 10.51\um} &\colhead{[NeII] 12.81\um} &\colhead{[NeV] 14.32\um} & \colhead{[NeIII] 15.56\um} & \colhead{[SIII] 18.71\um} & \colhead{[NeV] 24.32\um} & \colhead{[OIV] 25.89\um} & \colhead{[SIII] 33.48\um}& \colhead{PAH(6.2\um)}  &\colhead{[NeIII] 15.56\um}}

\startdata
NGC 4151&0.69,0.70 &1.0,1.0 &0.57,0.59 &1.51,1.54 &0.54,0.58 &0.50,0.51 &1.80,1.73 &0.49,0.49	& \nodata&	 .069\\
Markarian 3&0.60,0.60 &1.0,1.0 &0.64,0.65 &1.81,1.77 &0.55,0.56 &0.68,0.67 &2.14,2.14 &0.52,0.53 & \nodata&	 .123\\
I Zwicky 1	&$<$1.0,$<$0.7 &1.0,1.0 &$<$0.58,$<$0.56 &2.04,1.79&$<$0.4,$<$0.7 &$<$0.6,$<$0.4 & $<$0.9,$<$0.4 &$<$1.0,$<$0.8	& .010&	 .011\\
NGC 1275&$<$0.04,$<$0.03 &1.0,1.0 &$<$0.03,$<$0.03 &0.43,0.40 &0.15,0.21 &$<$0.04,$<$0.03 & 0.15, 0.19 & 0.25,0.27	& .020&	 .015\\
Centaurus A&0.071,0.071 &1.0,1.0 &0.12,0.12 &0.77,0.76 &0.23,0.24 &0.15,0.16 &0.71,0.65 &0.75,0.78 & .014&	 .065\\
NGC 7469&0.046,0.046 &1.0,1.0 &0.058,0.058 &0.17,0.17 &0.38,0.39 &0.074,0.073 &0.21,0.22 &0.52,0.46	& .208&	 .021\\
Markarian 231&$<$0.36,$<$0.39& 1.0,1.0 & $<$0.32,$<$0.35&$<$0.32,$<$0.29 &$<$0.38,$<$0.34 &$<$0.50,$<$0.55 &$<$0.55,$<$0.54 &$<$1.79,$<$1.72 &  .014&	 $<$.004\\
NGC 3079&$<$0.006,0.008 &1.0,1.0 &$<$0.009,0.01 &0.22,0.22 &0.12,0.12 &$<$0.015,$<$0.010 &0.088,0.090 &0.57,0.60	&.515&	 .053\\

\enddata
\tablenotetext{1}{Line intensities are relative to [NeII] 12.81\um with these relative intensities in each nod position in the high resolution listed separately; actual fluxes in this line from the average of two nod positions in units of 10$^{-20}$ W cm$^{-2}$ are 13.4 for NGC 4151, 10.0 for Markarian 3, 0.28 for I Zwicky 1, 4.80 for NGC 1275, 19.3 for Centaurus A, 20.0 for NGC 7469, 1.85 for Markarian 231, and 10.4 for NGC 3079.  ISO measurements for this line in the 6 objects in common are 11.8 for NGC 4151, 4.7 in Markarian 3, 0.65 in I Zwicky 1, 2.9 in NGC 1275, 22.1 in Centaurus A, and 22.6 in NGC 7469. For the diagnostic [OIV] 25.9\,\um/ [Ne II] 12.8\,\um, ISO-derived ratios are 1.7 for NGC 4151, 2.7 for Markarian 3, $<$ 0.9 for I Zwicky 1, $<$ 0.17 for NGC 1275, 0.44 for Centaurus A, and 0.14 for NGC 7469.} 
\label{tab3}
\end{deluxetable}
\clearpage









\end{document}